\documentclass[twocolumn]{IEEEtran} 

\usepackage{mymacros}
\usepackage{times,mathptm}
\usepackage[final]{graphicx}

\makeatletter

\long\def\@caption#1[#2]#3{\par\addcontentsline{\csname ext@#1\endcsname}{#1}{%
   \protect\numberline{\csname the#1\endcsname}{\ignorespaces #2}}
   \begingroup \@parboxrestore \normalsize
     \@makecaption{\csname fnum@#1\endcsname}{\ignorespaces #3}{#1}\par
   \endgroup}

\def\figurestring{figure}

\long\def\@makecaption#1#2#3{
    \vskip 5pt
    \setbox\@tempboxa\hbox{\small #1.~ #2}
    \ifdim \wd\@tempboxa >\hsize
       \setbox\@tempboxa\hbox{\small #1.~ }
       \setlength\captionindent{\wd\@tempboxa} \divide\captionindent by 2
       \parbox[t]{\hsize}{\small \hangindent \captionindent \hangafter=1%
       \unhbox\@tempboxa#2}
    \else \hbox to\hsize{\small\hfil\box\@tempboxa\hfil}
    \fi
}

\makeatother

\newlength{\mylen}

\graphicspath{{figures/}}

\newcommand{\note}[1]{\newline \emph{Note: #1}}

\begin{document}

\title{Performance of the Light Trigger System 
       in the Liquid Xenon Gamma-Ray Imaging Telescope LXeGRIT}

\author{Uwe~Oberlack, Elena~Aprile, Alessandro~Curioni, Karl-Ludwig~Giboni 
 \thanks{Columbia Astrophysics Laboratory, Columbia University, New York, 
         NY 10027, USA 
  \newline E-mail:
  \newline U.~Oberlack: oberlack@astro.columbia.edu
  \newline E.~Aprile: age@astro.columbia.edu
  \newline A.~Curioni: curioni@astro.columbia.edu
  \newline K.-L.~Giboni: kgiboni@astro.columbia.edu}
}

\markboth{IEEE Transactions On Nuclear Science --- Manuscript submitted November 29, 2000}
{U.~Oberlack et al.: Performance of the LXeGRIT Light Trigger System}
\maketitle

\begin{abstract}
LXeGRIT is a balloon-borne Compton telescope for \MeV\ gamma-ray astrophysics,
based on a liquid xenon time projection chamber with charge and light readout.
The energy and direction of an incident gamma-ray is reconstructed from the
three-dimensional locations and energy deposits of individual interactions
taking place in the homogeneous detector volume. While the charge signals
provide energy information and X-Y-positions, the fast xenon scintillation light
signal is used to trigger the detector. The drift time measurement, referred to
the time of the trigger signal, gives the Z-position with the known drift
velocity.  The light is detected by four UV-sensitive photomultiplier tubes
(PMTs). The logical OR of the PMT signals triggers the data acquisition system
with an efficiency which depends on the event energy and location, as well as on
the discriminator thresholds used on the individual PMTs. Results from
experiments with a tagged \nuc{22}{Na} source give the spatial distribution of
the light trigger efficiency for 511~\keV\ gamma-rays. When averaged over the
whole sensitive volume and all PMTs, the trigger efficiency is 47\% or 40\% for
two discriminator windows used during the LXeGRIT balloon flight of 1999. These
values are strongly affected by the different sensitivity of each PMT. The
corresponding average efficiency at 511~\keV\ for the best of the four PMTs is
in fact 63\%, and approaches 100\% for interactions taking place in a small
volume right above the PMT.
\end{abstract}



\section{Introduction}
Gamma-rays in the MeV energy band have a great scientific potential for
astrophysics. They directly address topics as diverse as
nucleosynthesis, supernova mechanisms, star formation and distribution of
massive stars, or the physics of accreting black holes \cite{GRAPWG:99}. Yet,
imaging of cosmic MeV gamma-rays is notoriously
difficult due to the lack of focussing optics, a broad minimum in the cross
section of photons with matter, and small source fluxes. This results in
the requirement of large-volume, massive detectors and long observing times. The
situation is further complicated by the atmosphere's opaqueness for gamma-rays,
requiring telescopes to be put into space or near-space where intense radiation
fields generate high background levels and thus signal-to-background ratios of
typically 1:100 or less.  To meet the demanding sensitivity requirements of
current astrophysical questions, new detectors are needed which combine high
efficiency, great background suppression capabilities, and a large
field-of-view. A Compton telescope for is considered the most
promising instrument design for MeV gamma-ray astrophysics,
following the pioneering achievements of the first
and only Compton telescope in space to-date,
CGRO/COMPTEL\cite{VSchoenfelder:00:sources}.

\begin{figure}
\includegraphics[width=\linewidth,clip]{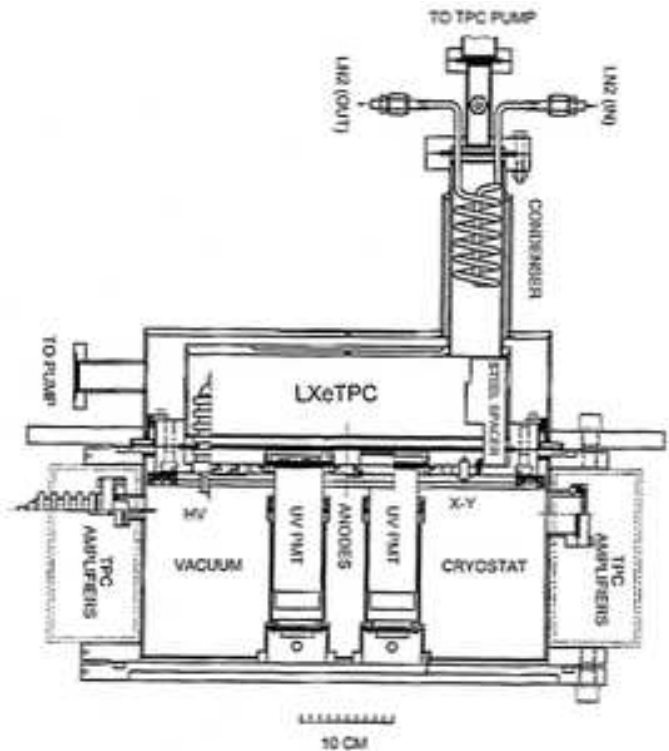}
\caption{\label{f:tpc:drawing} Mechanical design of the LXeTPC}
\end{figure}
LXeGRIT is a new type of Compton telescope in which gamma-rays are imaged in a
liquid xenon time projection chamber (LXeTPC), which combines charge and light
readout for three-dimensional localization and energy measurement in one
homogeneous volume.  A simplified mechanical drawing of the LXeGRIT detector is
shown in Fig.~\ref{f:tpc:drawing}. The TPC structure, with a sensitive area of
$20 \times 20 \textrm{ cm}^2$, and a drift region of 7 cm, is shown in the photograph of
Fig.~\ref{f:tpc:wirestruc}, where the cathode plate has been removed to view the assembly of
grid,X-Y wires and anodes. The charge clouds produced by gamma-ray interactions in the liquid,
are drifted in an electric field of  1~kV/cm, kept homogeneous by a series of shaping rings.
The signal induced by the drifting ionization electron clouds are detected by two arrays of
orthogonal wires (62-X and 62-Y), providing the two-dimensional localization of an interaction. 
The depth of the interaction
(Z-position) is inferred from the drift time measurement, referred to the event 
time zero provided by the scintillation signal, also produced in an interaction. 
The energy deposited in each interaction is measured from the collected charge signal(s)  on one of
four anodes. The induction and collection signals  are digitized by
flash ADCs with a sampling rate of 5~MHz and a resolution of 8 and 10~bit,
respectively \cite{EAprile:98:electronics}.

The TPC structure is mounted on four ceramic rods, screwed on the 16.5" stainless
steel flange, on which HV and signal feedtroughs are welded along with four
quartz windows, one under each of the anode meshes, to allow light transmission
to the PMTs. The flange closes the detector in a
cylindrical vessel, filled with about 8 liters of high purity liquid xenon. A vacuum
cryostat provides the thermal insulation. The liquid temperature is maintained
at $\sim -95\deg$~C within a few degrees, by controlling the vapor pressure
with a flow of liquid nitrogen through a condenser.  The detector, originally developed as a
laboratory prototype to test the feasibility 
of a large volume
LXeTPC and to demonstrate its performance with gamma-rays, was later converted to a balloon payload for observations of cosmic
gamma-ray sources. The first balloon flight of LXeGRIT, in 1997, proved the
reliable operation of the TPC in  near-space environment, validating the design
of various critical mechanical, electrical, and cryogenics systems. Shortcomings
associated with the data acquisition system (DAQ) and the trigger electronics,
identified during this flight, have been resolved with the development of a
new trigger processing unit and a new DAQ processor and software \cite{EAprile:01:daq}.
LXeGRIT, with the upgraded systems, performed successfully during two long
balloon flights in 1999 and 2000. A description of the payload in its 1999
flight configuration and results from calibration and flight data are given in
\cite{EAprile:00:spie00:performance,EAprile:00:spie00:flight99}.

\section{The LXeGRIT Light Trigger System}
\begin{figure}
\includegraphics[bb=43 60 334 286,width=\linewidth,clip]{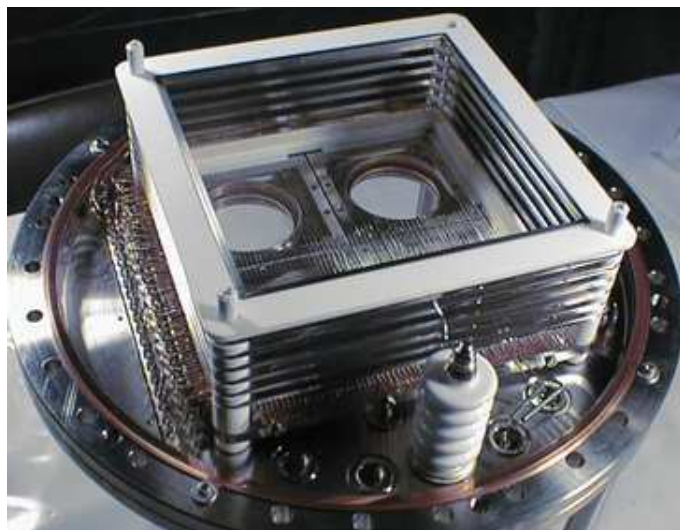}
\caption{\label{f:tpc:wirestruc} A photo of the assembled TPC structure taken with the cathode 
removed to show the readout structure and the windows through
which the PMTs view the LXe.}
\end{figure}

Liquid xenon is an excellent scintillator with a high photon yield similar to
that of NaI (Tl) but with a much shorter decay time, consisting of a fast ($<
5$~ns) and a slow (27~ns) component \cite{SKubota:79:free:elec,EAprile:90:ITNS}.
The scintillation emission peaks in the ultraviolet at 175~nm. 
In the LXeGRIT TPC, 
the scintillation light is detected by four 2" UV-sensitive PMTs (Electron Tubes 9813QA),
coupled to the liquid vessel through four quartz windows, 2.4" in 
diameter. Two of the windows, mounted on the detector's flange,
are visible in the photograph of Fig.~\ref{f:tpc:wirestruc}. 
A liquid xenon layer of about 3 cm separates the anode meshes from the
windows. Interactions in this layer produce a light signal without a charge signal and
thus contribute background triggers.  
The PMTs are mounted in the vacuum
cryostat, as shown in Fig.~\ref{f:tpc:drawing}, at a few millimeters from the
windows. The typical quantum efficiency of the PMT photocathode is about 15\%.

While the entire sensitive volume is visible to the PMTs, this TPC
design is not optimal for light collection.  Since the cathode and the field
shaping rings around the drift region have little reflectivity in the UV, solid
angle effects are expected to be significant. The W-value for scintillation
light in liquid xenon is similar to that for charge, resulting in a large
photon yield, i.e., $\sim 6 \cdot 10^4$ photons are generated by a
$\gamma$-interaction of 1~\MeV. However, the number of photons that reach a PMT
is drastically reduced by several factors. The largest loss is due to the small
solid angle of each PMT for most interaction locations. This results in a
typical reduction factor of $\sim 0.02$, varying widely across the
chamber. Further losses include quenching of the recombination scintillation
by the electric field (on the order of $0.5$), optical transmission
of the grid, X-Y wires, and anode mesh, losses at the LXe/quartz interface,
optical transmission of the quartz window and reflection at the PMT window.
The
resulting number of photons is too small for good energy resolution. The light
signal is dominated by solid angle effects, yet it is sufficient to trigger the TPC down to gamma-ray energies as low as 
100 keV. 
For the 1997 and 1999 balloon flights,
the PMTs were operated at negative high voltage (HV). To minimize noise from capacitive coupling 
to the input of the highly sensitive anode preamplifiers, positive HV was used instead for the 2000 flight. 
The PMT signals are processed by
charge-sensitive preamplifiers (Clear Pulse Model 5016 Dual CSA) and fed into a custom-made
trigger electronics unit (Clear Pulse Model 8823A), where the signals are
discriminated and logically combined. The low threshold of a discriminator window is used to suppress 
noise pulses and to select a minimum signal amplitude. The high threshold can be selected to 
ignore high-energy signals from cosmic rays. The trigger signal,
derived from the logical ``OR'' of the four PMTs, can be vetoed by a coincidence
signal of plastic and NaI scintillators, used to shield the TPC during 
the 1997 and 1999 flights. For the 2000 flight, all shields were removed, including the plastic counters. For any PMT signal
above the low discriminator threshold, the trigger logic also checks
whether a trigger signal was generated  within the previous 50~\us, in which case it issues an abort
signal to reset the data acquisition system. The rates after each step in the logic are counted by a 16 channels
scaler unit (Clear Pulse Model 8823B), and read by the data
acquisition processor via a VME interface. For further details on the LXeGRIT trigger system and data
acquisition system we refer to \cite{EAprile:01:daq}.


\section{Set-up for the Trigger Efficiency Measurement}

\begin{figure}
\includegraphics[width=\linewidth,clip]{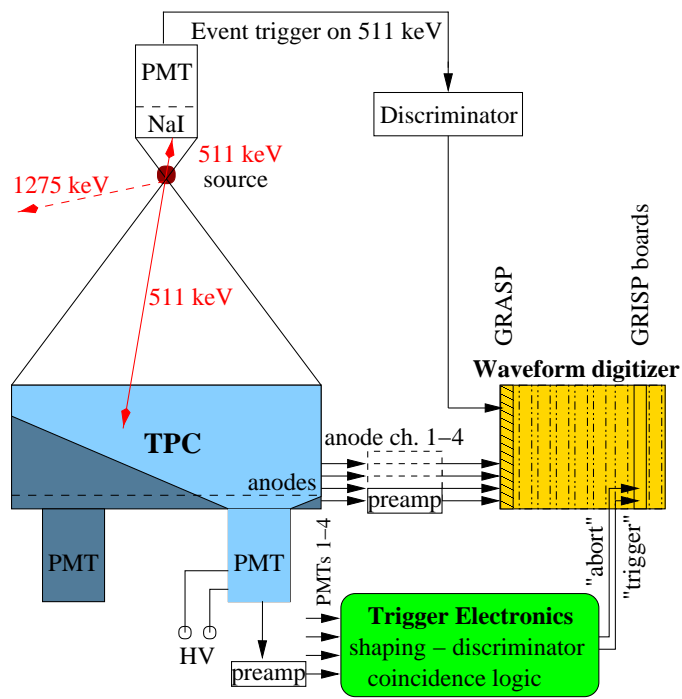}
\caption{\label{f:tagged:setup} Set-up of the spatially resolved trigger
efficiency measurement using the LXeGRIT flight electronics and a tagged
\nuc{22}{Na} source.}
\end{figure}

\begin{table}
\centering
\caption{\label{t:settings} High voltage and trigger threshold settings.}
\tabcolsep 1ex
\begin{tabular}{cccccc}
Setting & PMT 1  & PMT 2  & PMT 3  & PMT 4  & Window \\ 
\hline
 A  & -1870 V & -1800 V & -1850 V & -1870 V & 10--128 \\
 B    & -1870 V & -1800 V & -1850 V & -1870 V & 10--255 \\
\end{tabular}
\end{table}

The trigger efficiency of the LXeGRIT instrument was measured in a dedicated
experiment with a tagged \nuc{22}{Na} source.  The HV on the four PMTs of the
light readout, and the discriminator low and high thresholds were the same as
during the 1999 flight (see Table~\ref{t:settings}). Most of the flight data
were taken with the 10--128 window, on all four PMTs.  The TPC three-dimensional
imaging capability provides the unique opportunity not only to determine the
overall trigger efficiency, but also its spatial distribution within the
sensitive volume. The trigger efficiency as a function of the energy deposited
in each volume element can be directly applied to Monte Carlo simulations, in
the broader context of representing the overall instrument response.

The
FADC readout of the charge signals on wires and anodes provides the energy deposit
and the location of each gamma-ray interaction of an event. To initiate
the recording of the digital image of the event, an external trigger is
required. In the trigger efficiency experiment, one of the two back-to-back 511
keV gamma-rays from a \nuc{22}{Na} source is detected by a NaI(Tl) counter. 
This signal triggers the TPC.  Fig.~\ref{f:tagged:setup} shows the experimental
set-up. The \nuc{22}{Na} source is centered on the TPC and viewed by a NaI(Tl)
counter from the top. The distances between detector and source and between
source and  counter are adjusted such that for almost each 511~\keV\ photon
detected by the NaI(Tl), the other 511~\keV\ photon will hit the TPC,
covering the entire sensitive volume. To select annihilation photons with high
efficiency, we restrict the NaI(Tl) trigger to the 511~\keV\ photopeak with
discriminators.  The Xe scintillation signals from the four PMTs were treated
with the LXeGRIT trigger electronics. Measurements were made with either one PMT
at the time or all four PMTs, connected to the input of the trigger logic unit.  The
single-PMT measurement (HV on the other three PMTs was turned off)
was done to determine the
response of the individual PMTs and their contribution to the trigger
efficiency, while the measurement with all four PMTs provided the overall
efficiency map.

The `trigger' and `abort' signals from the trigger logic unit were
fed into two spare channels of the TPC digitizers, so as to record the occurance of a
light trigger on an event-by-event basis. The spatially resolved trigger
efficiency is derived from the ratio of events with `trigger' signal (and
without `abort' signal) divided by all events in the same spatial bin and energy window.  
In this
set-up, the trigger efficiency is measured only for energy deposits up to
511~\keV.

\section{Results}

\begin{figure}
\centering
\setlength{\mylen}{5cm}
\includegraphics[bb=79 296 306 485,width=\linewidth,clip]
                {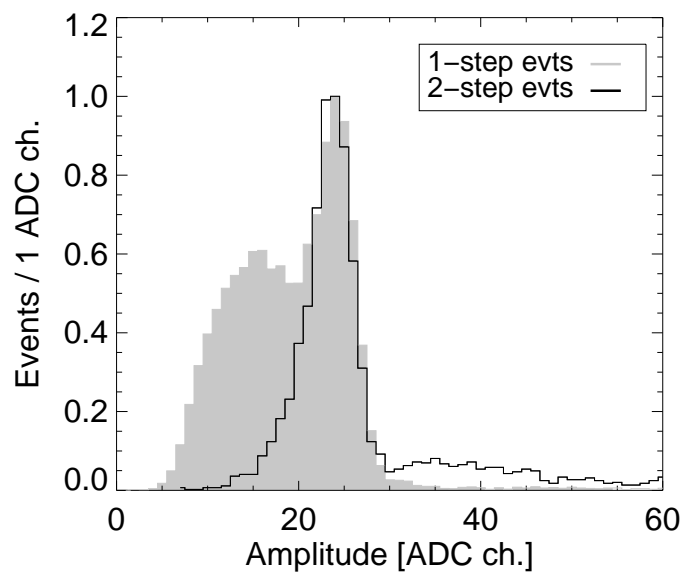}
\caption{\label{f:tagged:spectra} Normalized \nuc{22}{Na} spectra from single-
and two interaction events measured with the tagged-source set-up, for all
triggers. The amplitudes were corrected for gain differences among the anodes
and for charge loss due to electron attachment.
}
\end{figure}

\begin{figure*}[tb]
\centering
\setlength{\mylen}{0.4\linewidth}
\includegraphics[bb=77 514 310 715,width=\mylen,clip]
                {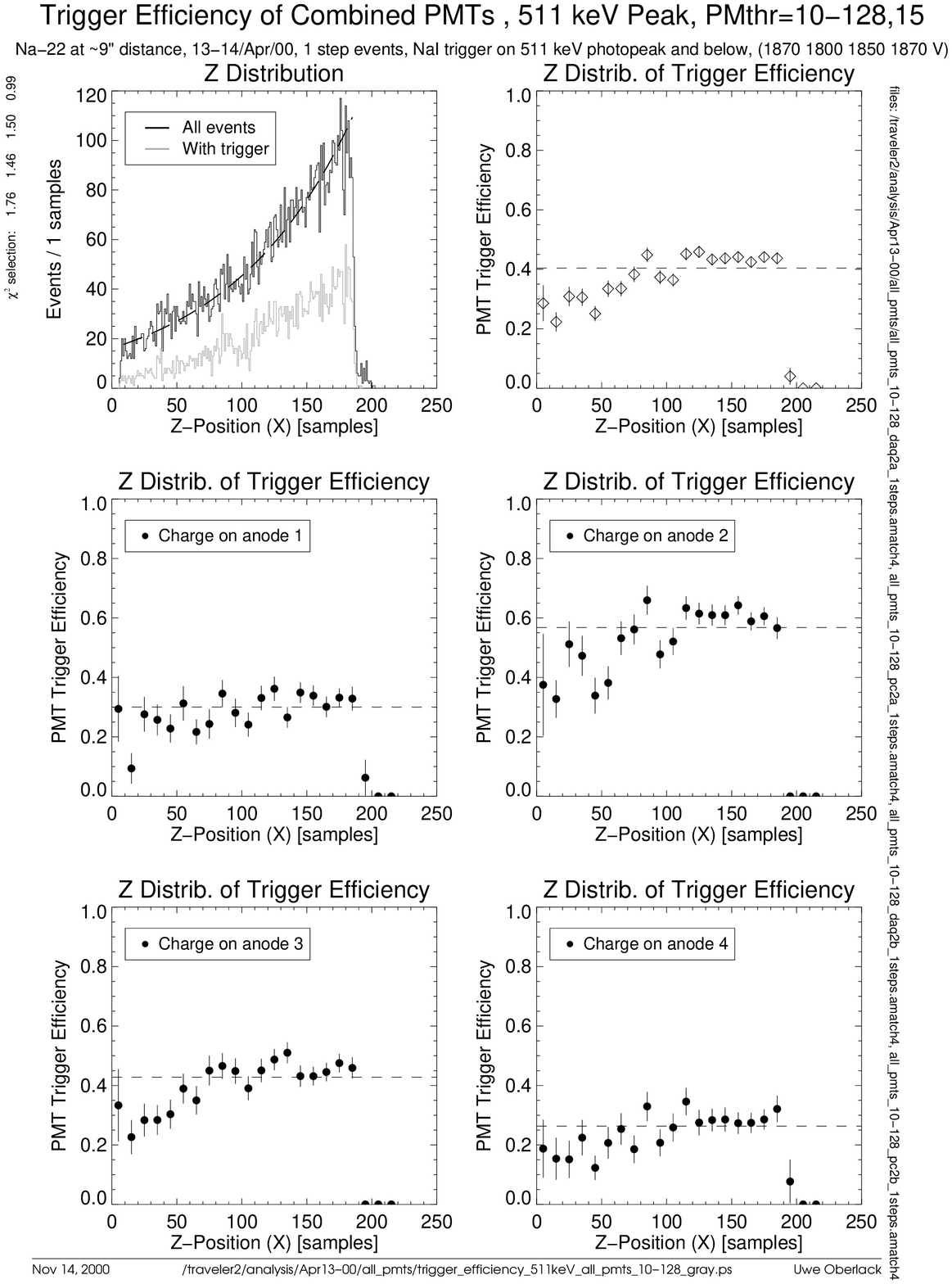}
\includegraphics[bb=77 514 310 715,width=\mylen,clip]
                {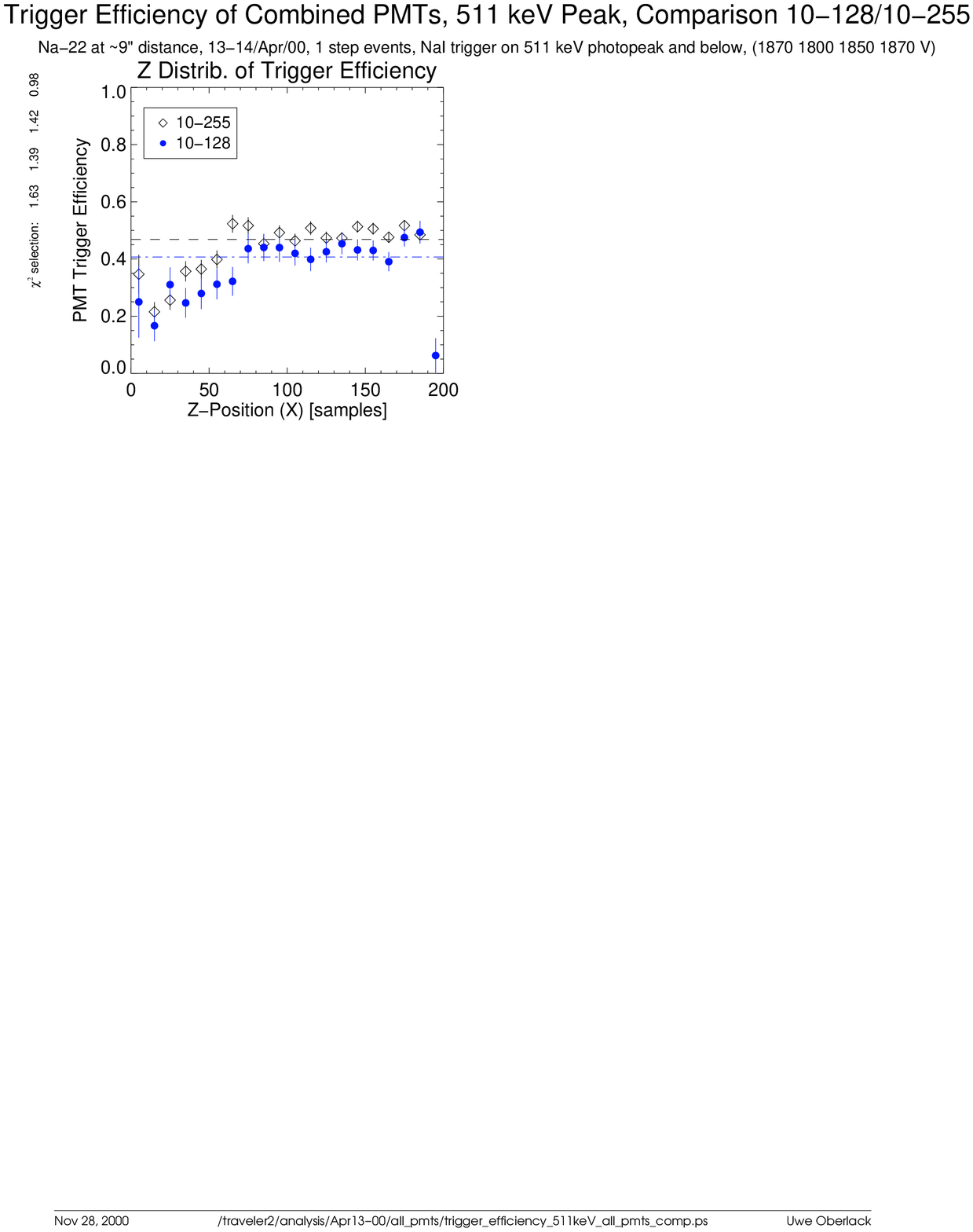}
\caption{\label{f:zdist:511:F99:all} $Z$-distributions for 511~\keV\ 1-step
events with HV and threshold settings as during the 1999 ballon flight and with
all four PMTs connected.
Left: $Z$-distribution of all events and of events with a trigger signal.
Right: $Z$-dependence of the overall trigger efficiency, integrated over all
$X,Y$, for two discriminator windows (setting ~A and ~B)}
\end{figure*}

\begin{figure*}[tb]
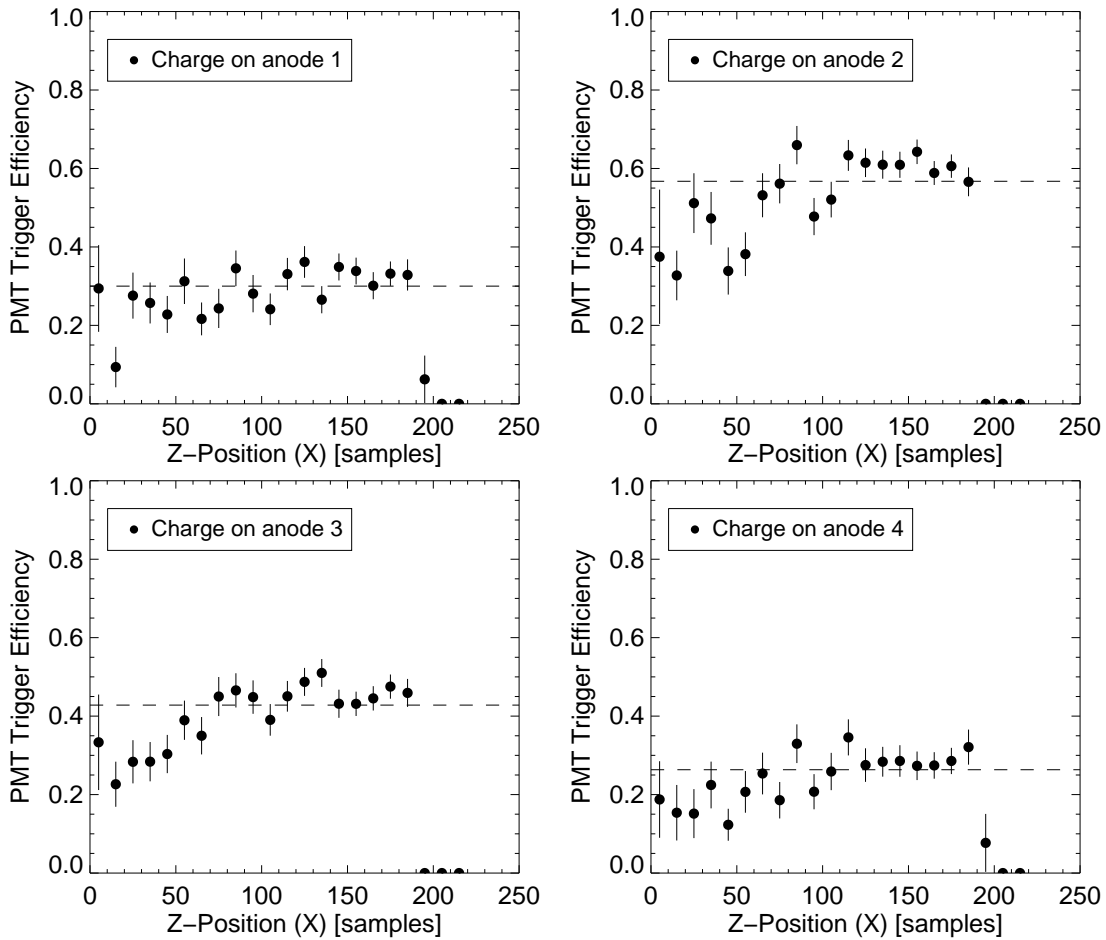

\centering
\includegraphics[bb=77 283 550 483,width=0.8\linewidth,clip]
                {trigger_efficiency_511keV_all_pmts_10-128_p1.ps}
\\
\includegraphics[bb=77 52 550 252,width=0.8\linewidth,clip]
                {trigger_efficiency_511keV_all_pmts_10-128_p1.ps}
\caption{\label{f:zdist:511:F99:single} $Z$-distributions of the trigger
efficiency for 511~\keV\ 1-step events, selected by the
charge deposit on individual anodes, which mainly reflects triggers on the
corresponding PMTs. The discriminator window was 10--128 (setting ~A). 
}
\end{figure*}

Fig.~\ref{f:tagged:spectra} shows the \nuc{22}{Na} energy spectrum obtained from the digitized
anode signals, for events recognized with either one interaction (1-step events)
or two interactions (2-step events), regardless of a xenon light trigger.  The
amplitudes have been corrected for gain 
differences among the four anodes and for charge loss due to electron
attachment. The electron lifetime is derived from a fit of the $Z$-dependence of
the photopeak position in the 1-step events, resulting in a best-fit value of
332~\us, with a systematic uncertainty of about $\pm 50~\us$. The 1-step events distribution show
a strong Compton tail, which is largely due to backscattered Compton events at
high $Z$. Very few events show energies beyond the 511~\keV\ photopeak, centered
at \mbox{24.1$\pm 0.2$ ADC} channels, indicating that the contamination of the sample
with background is small. The coincident 1275~\keV\ line does not appear in this
spectrum, since, due to the geometry of the set-up, almost every event will
produce at least one step from a 511~\keV\ photon in the detector. Events from
1275~\keV\ gamma-rays would therefore have to show at least two steps. Indeed,
the 2-step distribution shows significantly more events beyond the 511~\keV\ line. It
also shows the effect of Compton tail suppression with multiple interaction
events.  On-line event selections, such as minimum energy thresholds on wire and
anode signals, as well as a minimum and maximum number of wire hits, also affect
the shape of the distribution below the annihilation line. In both spectra, the
511~\keV\ line stands out clearly, with similar peak position and width.

\begin{figure*}
\centering
\setlength{\mylen}{4.6cm}
\includegraphics[bb=319 524 552 729,height=\mylen,clip]
                {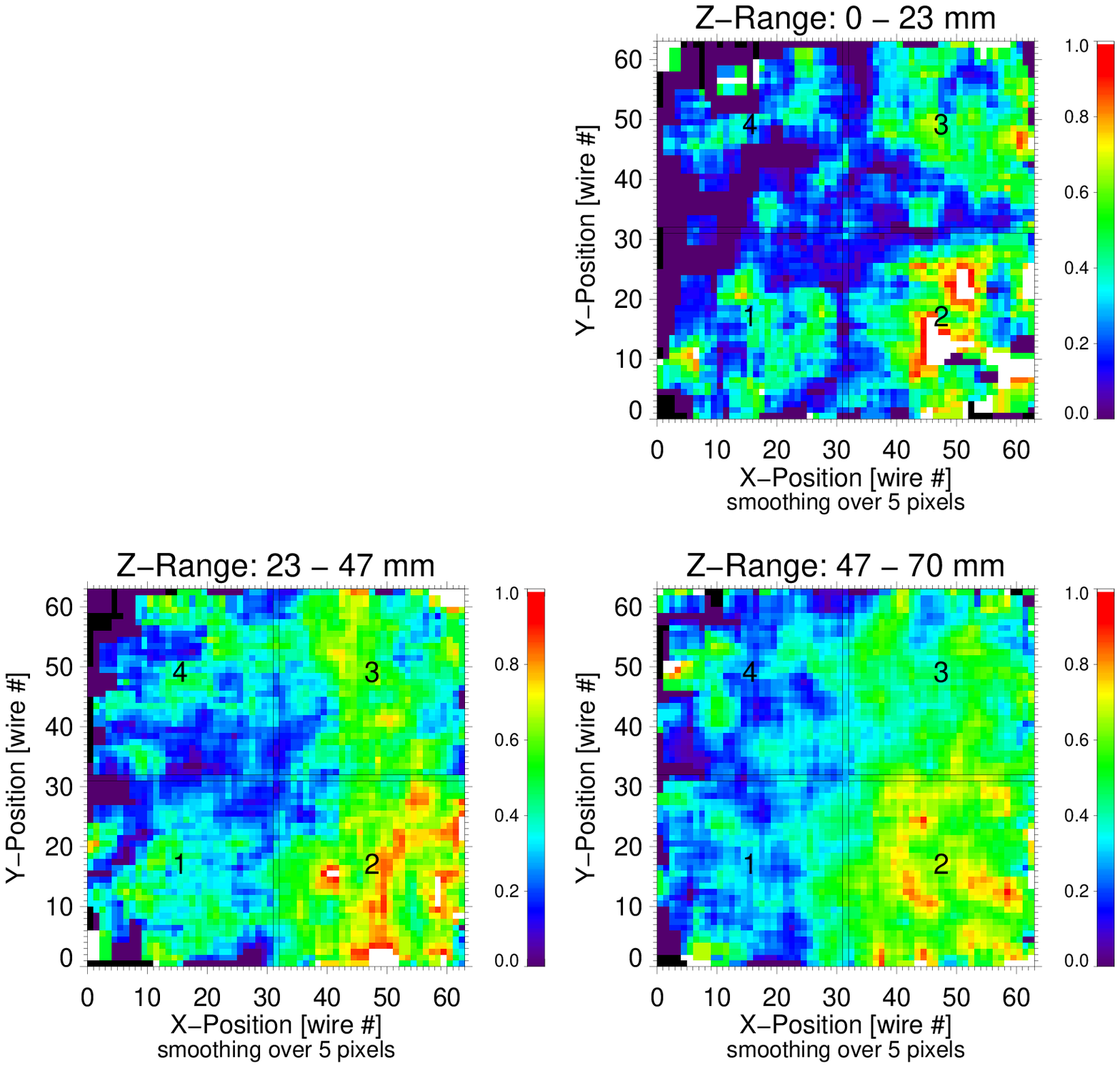}
\includegraphics[bb=79 293 552 498,height=\mylen,clip]
                {trigger_efficiency_511keV_all_pmts_comp_p9.ps}
\\[1ex]
\includegraphics[bb=319 524 552 729,height=\mylen,clip]
                {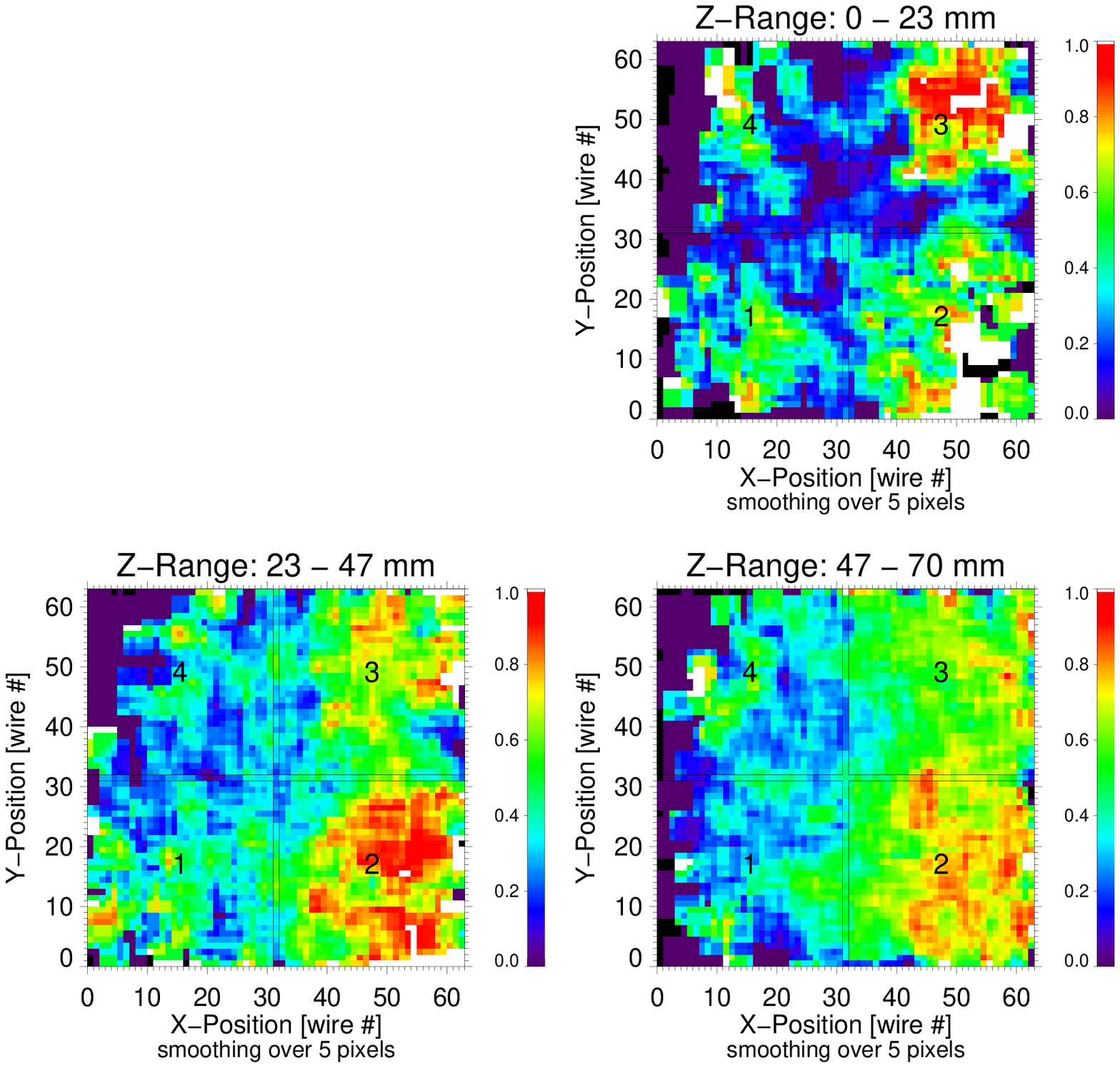}
\includegraphics[bb=79 293 552 498,height=\mylen,clip]
                {trigger_efficiency_511keV_all_pmts_comp_p3.ps}
\caption{\label{f:xydist:511:all} 
$(X,Y)$-distributions of the trigger efficiency for 511~\keV\ 1-step events
for three different depths of the detector. The four anodes are
labeled. The top panel shows data with setting~A (10--128) and the lower panel
with setting~B (10--255), with all four PMTs
connected. The images are smoothed over 5 pixels, and black pixels indicate no
data. 
}
\end{figure*}
\begin{figure*}
\centering
\setlength{\mylen}{4.6cm}
\includegraphics[bb=319 293 552 498,height=\mylen,clip]
                {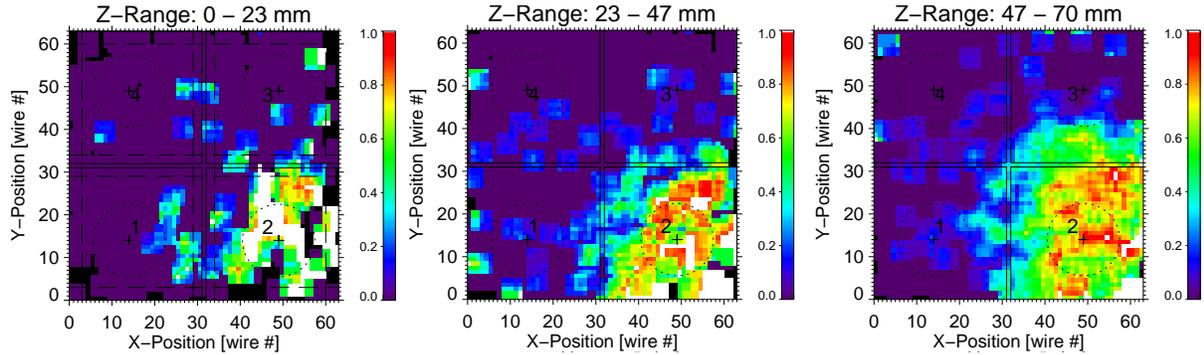}
\includegraphics[bb=79 61 552 266,height=\mylen,clip]
                {trigger_efficiency_pmt2_10-255_p3.ps}
\caption{\label{f:xydist:511:pmt2} $(X,Y)$-distributions of the trigger
efficiency as in Fig.~\ref{f:xydist:511:all}, but for PMT~2 only and with
setting~B. In the display of the lower third of the detector, dashed lines
indicate the anode frame and the 2" PMTs.
}
\end{figure*}

Fig.~\ref{f:zdist:511:F99:all} shows the $Z$-distribution for all 511~\keV\
1-step events, and the corresponding one for events with a light trigger.  All
four PMTs were connected with the HV as in Table~\ref{t:settings} and for the
discriminator setting~A (10--128). The distribution is well fitted by an
exponential, consistent with the expected attenuation length of about 4~cm for
511~\keV\ gamma-rays in LXe.  From this data and the corresponding data taken
with setting~B (10--255), we calculate the light trigger efficiency averaged
over all $X$ and $Y$ (thus over all four PMTs), as a function of $Z$. This is
shown on the right panel of the figure, for both settings. The efficiency
increases with increasing $Z$, resulting in average numbers of 40\% and 47\%,
respectively. Several effects influence the $Z$-distribution of the trigger
efficiency:
\begin{enumerate}
\item On average over the chamber, the solid angle of each PMT decreases with
      increasing $Z$. 
\item The area seen by each PMT increases however with $Z$ due to the shadowing effect 
      of the metal seal around the window and of the frame that holds the anode
      mesh.
\item The upper trigger threshold cuts into events predominantly at low $Z$,
      where the larger PMT solid angle results in larger light signals.
\item At low $Z$, reflection losses increase because of the higher index of
      refraction of quartz compared to the vacuum on top of the PMT
      photocathode, which leads to total reflection for large
      incident angles.
\end{enumerate}

The overall trigger efficiency is strongly affected by the different sensitivities of
the four PMTs. This is shown in Fig.~\ref{f:zdist:511:F99:single} where the data
from the measurement with setting~A have been subdivided into the four volumes on top of the
individual anodes. The four $Z$-distributions reflect mainly the response of the
individual PMTs, which differ widely in noise levels and
therefore sensitivities. PMTs~2 and 3 are most sensitive. Events at
high $Z$ still generate triggers with high efficiency despite the reduced solid
angle and therefore reduced light signals. At \mbox{$Z$ \lsim\ 100} samples
(1~sample is 0.2~\us\, or about 0.4~mm), 
the effect of increasing viewing area 
dominates, and for setting~A, the effect of the upper threshold cut becomes
important at low $Z$. For PMTs~1 and 4, more events are lost below the lower
threshold at high $Z$, which over-compensates the effect of increasing area. For
PMT~1, the upper trigger threshold of 128 starts to cut into the distribution at
$Z < 40$~samples. This effect is weaker for PMT~4.

Fig.~\ref{f:xydist:511:all} shows the $(X,Y)$-distribution of the trigger
efficiency for 511~\keV\ 1-step events for three different $Z$-slices of the
detector volume. The upper three images are for setting~A and the lower ones for
setting~B, with the increased upper discriminator threshold.  The effect of
a larger area seen by the PMTs at high $Z$ (especially towards the center of
the detector) is apparent. The impact of the upper threshold cut is also
clearly seen, especially on PMTs~2 and 3, where the effect extends even to the
highest $Z$-positions. With setting~B, the trigger efficiency
approaches 100\% at low $Z$ in the areas centered on PMTs 2 and 3. 

The spatial distribution of the trigger efficiency of individual PMTs is 
more clearly seen from measurements with a single PMT. For
PMT~2 and setting~B, results are shown in Fig.~\ref{f:xydist:511:pmt2}. 
The trigger efficiency for 511~\keV\ events within an area of 2" diameter on top
of the PMT, indicated by the dashed circle, is about 75\%, averaged over all
$Z$. For setting~A, this efficiency drops to $\sim 60\%$. 

\begin{figure}[tb]
\centering
\setlength{\mylen}{5cm}
\includegraphics[bb=75 515 311 728,width=\linewidth,clip]
                {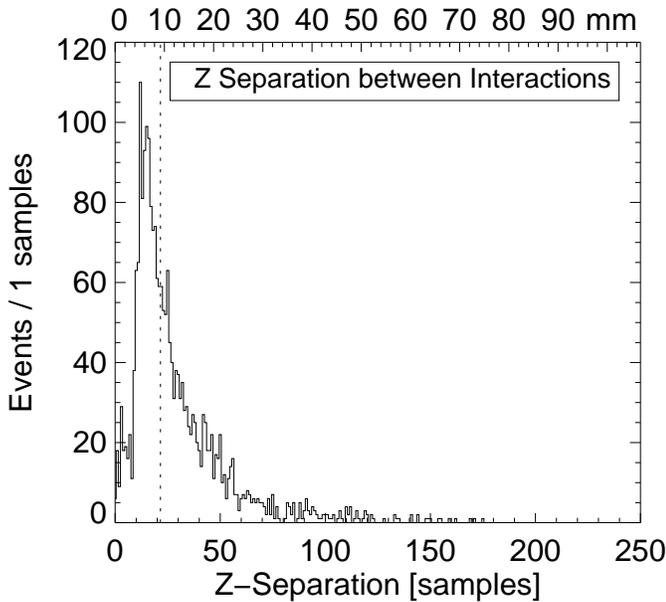}
\caption{\label{f:z-separation} $Z$-separation between two steps in the LXeTPC
for photons in the 511~\keV\ photopeak. The median of the distribution,
indicated by the dashed line, is at about 4~\us\ (20 samples) drift time, or
about 9~mm.
}
\end{figure}
Since LXeGRIT requires multiple-step events for Compton imaging, we are
interested in the trigger efficiency for this class of events. In principle,
this can be fully derived from 1-step events, if a complete trigger efficiency
map in energy and space were measured, and then applied to Monte Carlo
simulations. On the other hand, we can also directly measure the approximate spatial
distribution of the trigger efficiency for multiple-step events from the
tagged-source experiment, even though it
is unknown which interaction caused the trigger signal. We studied tthe spatial
distribution for 2-step events, calculating the efficiency either by counting
each position separately, or by weighting the $Z$-positions by energy deposit.

\begin{figure}[tb]
\centering
\includegraphics[bb=318 515 550 715,width=\linewidth,clip]
                {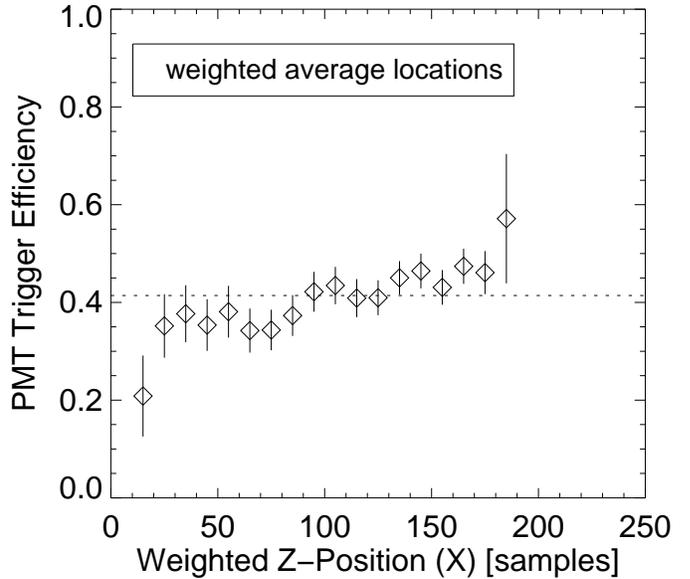}
\caption{\label{f:zdist:twosteps} $Z$-distribution of the trigger efficiency for
two-step events in the LXeTPC, for photons in the 511~\keV\ photopeak. We
compute the amplitude-weighted average as effective $Z$-position. The result is
very similar to the result for 1-step events.
}
\end{figure}

\begin{figure*}[tb]
\setlength{\mylen}{0.48\linewidth}
\includegraphics[bb=310 526 522 732,width=\mylen,clip]{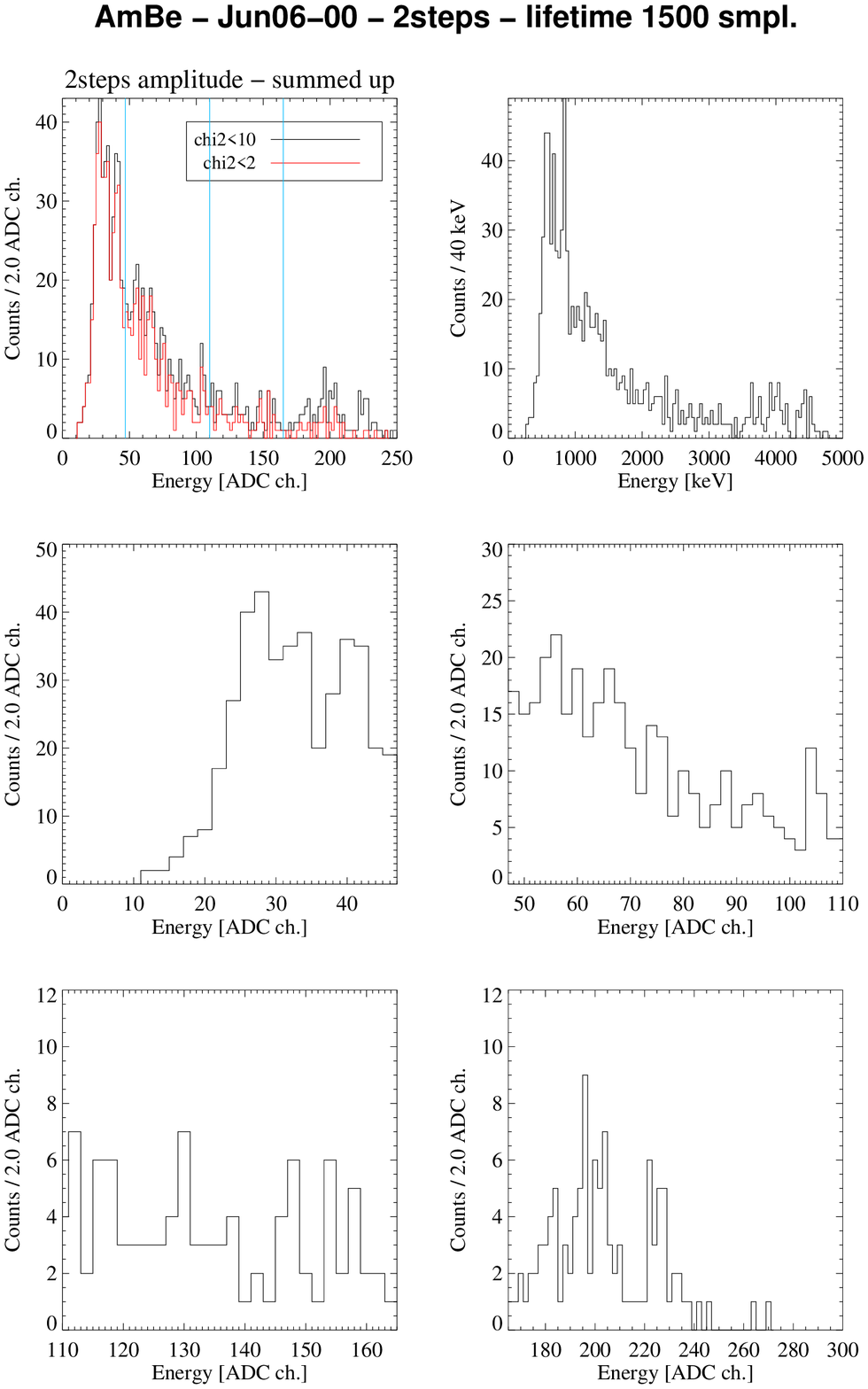}
\hfill
\includegraphics[bb=310 526 522 732,width=\mylen,clip]{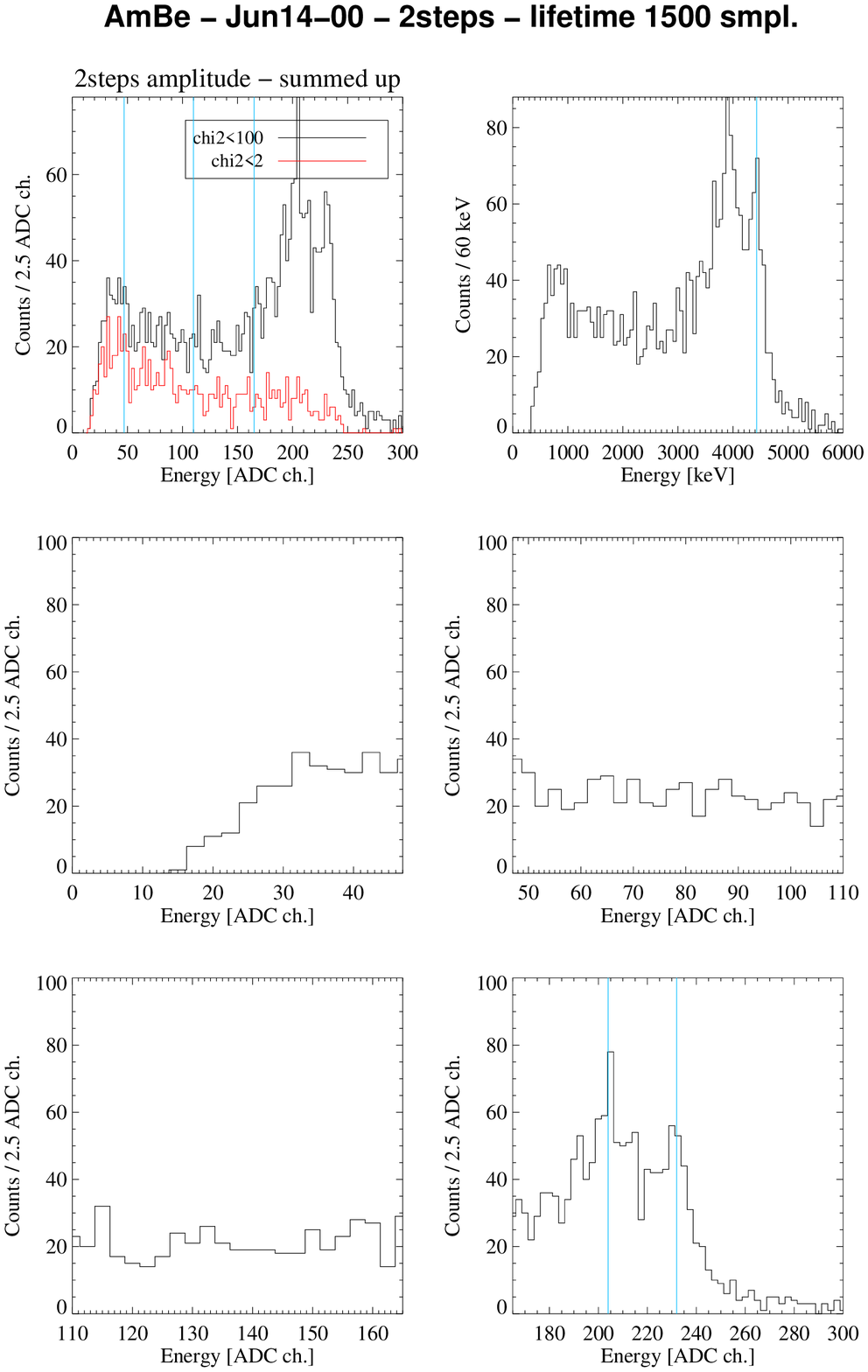}
\caption{\label{f:ambe:comparison} Am-Be spectra in flight 1999 settings
      (left) and with reduced HV settings and open upper
      threshold (right).}
\end{figure*}

Fig.~\ref{f:z-separation} shows the $Z$-separation between two interactions for
511~\keV\ \g-rays.  The dip in the distribution below about 10~samples results
from the loss of events due to the inability to fit the amplitude of the two
steps on one anode, when their $Z$-separation is so small.  Since the median
separation in $Z$ is only $\sim 9$~mm, we expect that the spatial distribution
of the trigger efficiency for these events be similar to that for single
interaction events. This is indeed the case, as shown in
Fig.~\ref{f:zdist:twosteps}, for setting~A. At higher energies, the spatial
separation between interactions becomes larger and therefore events spread over
a larger fraction of the sensitive volume. In this case, the spatial
inhomogeneities of the trigger efficiency become less relevant.

Results from these experiments and from the LXeGRIT 1999 balloon flight data
\cite{EAprile:00:spie00:flight99}, have led us to revise the PMT HV values and
the discriminator thresholds in order to reduce the noise and the trigger
sensitivity at low energy in order to improve the instrument response at \MeV\
energies. In preparation for the Fall 2000 balloon flight of LXeGRIT, the PMT
amplifiers and logic unit were also modified.  The combination of lower HV,
shorter shaping time, and open discriminator window resulted in a dramatic
impact of the instrument trigger response to higher energy photons. This is
demonstrated by the energy spectra of an AmBe source from 2-step events in
Fig.~\ref{f:ambe:comparison}. The spectrum to the left was taken with setting~A
as in 1999 flight configuration. It shows a strong enhancement of low-energy
triggers, with the 4.43~\MeV\ photopeak largely suppressed. The spectrum to the
right was instead derived from a measurement where the PMTs HV was reduced by
several hundreds Volts and the upper discriminator threshold was turned off.
The picture is completely changed. The photopeak and the single escape peak are
now the dominant features in the spectrum.

\section{Summary}
We have presented results from experiments dedicated to a spatially resolved
measurement of the light trigger efficiency in the LXeGRIT TPC at 511~\keV.
Data were taken with the HV values and with the two window discriminator
thresholds used in the 1999 balloon flight. For the discriminator window 10--128
with which most of the flight data were acquired, the trigger efficiency,
averaged over the entire sensitive volume and all four PMTs, was $\sim 40\%$,
and $\sim 47\%$ for the 10--255 window. The large variation in performance of
the four PMTs and the higher efficiency value of $\sim 63\%$ on PMT~2 for the
10--255 window indicates that even with the current TPC design a significantly
improved value can be achieved. The trigger efficiency is found to vary widely
across the detector due to the different PMT sensitivies and to the large
variation in solid angle, as well as to other geometrical effects.  These
variations and the absence of UV reflectors result in poor light collection
efficiency and prohibit to set a precise lower energy threshold.  This
deficiency cannot be solved without significant changes in the chamber itself.
For the 2000 flight, we opted to operate at a lower trigger efficiency in order
to reduce the trigger rate from photons below the sensitivity of the charge
readout of 150 -- 200~\keV.

Additional experiments will extend the 3D trigger efficiency measurement to
higher energies. This can be achieved by rotating the \nuc{22}{Na} source set-up
by 90\deg, again triggering on the 511~\keV\ photopeak. Thus, the second
collinear 511~\keV\ photon cannot enter the TPC, while the coincident 1275~\keV\
\g-ray from the deexcitation of \nuc{22}{Ne} is emitted isotropically, and can
therefore hit the TPC.  Other options include triggering on one of two
coincident \g-rays (\nuc{60}{Co},\nuc{88}{Y}) or tagging of
$\beta^-$-radioactive sources with prompt \g-decays (e.g.,\nuc{137}{Cs},
\nuc{60}{Co}) with small plastic scintillation counters to detect the fast
electron released in the decay.

\section*{Acknowledgments}
This work was supported by NASA grant NAG5-5108 to the Columbia Astrophysics
Laboratory.

\bibliographystyle{IEEEbib2}
\bibliography{mnemonic_short,myreferences}

\end{document}